# Violation of the Wiedemann-Franz Law for ultracold atomic gases

Michele Filippone[1], Frank Hekking[2,3], and Anna Minguzzi[2,3]

[1] *Dahlem Center for Complex Quantum Systems and Institut für Theoretische Physik, Freie Universität Berlin, Arnimallee 14, 14195 Berlin, Germany*
[2] *Université Grenoble Alpes, LPMMC, F-38000 Grenoble, France and*
[3] *CNRS, LPMMC, F-38000 Grenoble, France*

We study energy and particle transport for one-dimensional strongly interacting bosons through a single channel connecting two atomic reservoirs. We show the emergence of particle- and energy-current separation, leading to the violation of the Wiedemann-Franz law. As a consequence, we predict different time scales for the equilibration of temperature and particle imbalances between the reservoirs. Going beyond the linear spectrum approximation, we show the emergence of thermoelectric effects, which could be controlled by either tuning interactions or the temperature. Our results describe in a unified picture fermions in condensed matter devices and bosons in ultracold atom setups. We conclude discussing the effects of a controllable disorder.

The recent observations of mesoscopic transport in cold atomic gases [1] pave the way to the investigation of phase-coherent transport with bosons. This stimulates theoretical studies [2] complementary to those in metallic systems. Bosons are exempted from Pauli principle and it is now possible to establish how this intrinsic difference may affect energy and particle transport in a framework different from electronic devices.

A fundamental hallmark of electronic transport in mesoscopic metals is the Wiedemann-Franz (WF) law [3]. This law establishes that the ratio between thermal conductivity $\mathcal{K}$, electric conductivity $g$, and temperature $T$, known as the Lorenz number $L$, is a combination of universal constants $L_0$ given by

$$L \equiv \frac{\mathcal{K}}{gT} = \frac{\pi^2}{3}\left(\frac{k_B}{e}\right)^2 \equiv L_0. \qquad (1)$$

The Drude [4] and the Fermi liquid theory of transport [5, 6] provide a microscopic interpretation of this law: low energy quasi-particles carry both charge and energy. Indeed, deviations from the WF law Eq. (1) are then considered a signature of the breakdown of the quasi-particle character of low energy excitations. Deviations have been observed in high-Tc superconductors [7], close to phase transitions [8] and in one-dimensional (1D) strongly interacting channels [9].

In 1D, the screening of interactions is much less effective, leading to the failure of the one-body picture. Low-energy collective modes emerge as a complicated superposition of the elementary constituents. The Fermi liquid theory breaks down and must be replaced by an effective hydrodynamic approach, the Luttinger Liquid (LL) theory [10]. The linearization of the spectrum close to the Fermi surface allows to describe the low-energy excitations as a collection of non-interacting bosonic oscillators, characterized by the sound velocity $u$ and the interaction parameter $K$. The emergence of neutral collective modes, responsible for energy transport, distinct from the elementary constituents, carrying charge, leads to the violation of the WF law [11–14], *i.e.* the Lorenz number is different from $L_0$ and depends on the interaction parameter.

It is an open issue to establish to which extent the analog of WF law applies for neutral interacting 1D bosons, where one compares heat and mass – rather than charge – transport. In the limit of infinite interactions, hard-core bosons or Tonks-Girardeau (TG) gas have the same currents as those of a free Fermi gas, as follows from an exact mapping solution [15]. In this limit, corresponding to $K = 1$, the WF law holds. On the other hand, decreasing the interaction strength away from the TG regime, strongly interacting 1D bosons are described by a LL with $K > 1$ [16, 17], instead of $K \leq 1$ as for fermions with repulsive interactions.

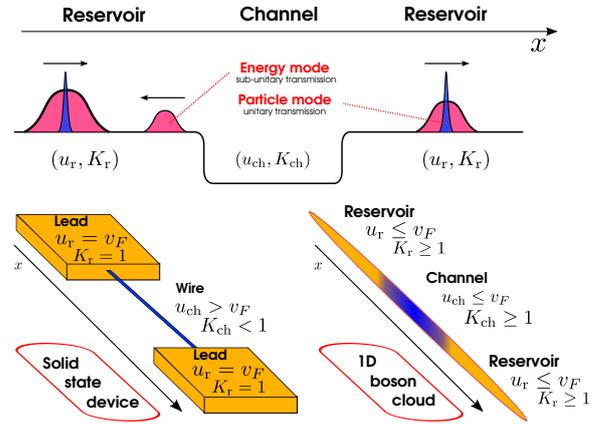

FIG. 1. Transport in inhomogeneous Luttinger liquids. Particle and energy transport are differently affected by the spatial variations of the LL parameters $(u, K)$. In solid state devices this can be observed in leads connected by a nanowire. In 1D bosonic clouds, the confining potential can be engineered in such a way to create two reservoirs collected by an atomic waveguide. Using ultracold atoms, it is possible to investigate a larger parameter region in which $K \neq 1$ not exclusively in the conducting channel.

In this Letter, we show that interacting 1D bosons, due to their strongly correlated nature, display a violation of the WF law. This has interesting consequences, relevant to 1D bosonic systems. For example, different time scales for the equilibration of temperature and particle imbalances between reservoirs emerge. These features could be directly accessed in a set-up similar as the one employed in recent experiments [1]. We will focus on the two-terminal setup sketched in Fig. 1. A left (L) and a right (R) reservoir are connected by a ballistic channel of length $d_{\rm ch}$. The interaction strength in the channel is assumed to be different from the one of the reservoirs, as could e.g. be engineered by adjusting the transverse confinement and hence the background density in each of the regions.

*Transport coefficients and model* - Particle and energy currents, $J$ and $J_{\rm E}$, generated by a difference of chemical potential $\Delta\mu = \mu_{\rm L} - \mu_{\rm R}$ or temperature $\Delta T = T_{\rm L} - T_{\rm R}$ are given by the transport matrix [18]

$$\begin{pmatrix} J \\ J_{\rm E} \end{pmatrix} = g \begin{pmatrix} 1 & s/T \\ s & L + s^2 \end{pmatrix} \begin{pmatrix} \Delta\mu \\ T\Delta T \end{pmatrix}. \qquad (2)$$

Its off-diagonal elements, related to Peltier and Seebeck effects, depend on the thermopower $s$, a manifestation of Onsager relations [19].

It is our aim to obtain and discuss all the elements of the transport matrix. For this purpose, we first derive an effective low-energy Hamiltonian governing the system depicted in Fig. 1. It is obtained by quantization of the classical equations of hydrodynamics in 1D [20]. These involve the continuity equation $\partial_t n + \partial_x(nv) = 0$ and Newton's law: $m n d_t v = -\partial_x P - n\partial_x V_{\rm ext}$. $n(x,t)$ and $v(x,t)$ are fields describing the gas density and velocity when submitted to the pressure $P$ and an external potential $V_{\rm ext}$. These equations are linearized close to equilibrium: we assume $n(x,t) = n_0(x) + \delta n(x,t)$ and introduce a first order displacement field $\vartheta$ such that $v(x,t) = \partial_t \vartheta(x,t)$. It is also useful to introduce the chemical potential $\mu[n]$ such that $\partial_x P = n\partial_x \mu[n]$. This encodes the information about interactions among the bosons. The linearized equations of motion [21] lead to the conserved quantity $\varepsilon = \int dx \{ \frac{mn_0}{2}(\partial_t \vartheta)^2 + \frac{1}{2}\frac{\delta\mu}{\delta n}\big|_{n_0} [\partial_x(n_0\vartheta)]^2 \}$, the energy of the system. We introduce the conjugate fields $\theta(x) = \pi n_0(x)\vartheta(x)$ and $\Pi(x) = m\partial_t\vartheta(x)/\pi\hbar$, describing density and current fluctuations. The standard quantization procedure $[\theta(x), \Pi(x')] = i\delta(x-x')$, with $u(x)K(x) = \pi n_0(x)\hbar/m$ and $u(x)/K(x) = \frac{1}{\pi\hbar}\frac{\delta\mu}{\delta n}\big|_{n_0}$ leads to the inhomogeneous LL model, thereby recovering [22–24]

$$\mathcal{H}_0 = \frac{\hbar}{2\pi}\int dx \left[ u(x)K(x)(\pi\Pi)^2 + \frac{u(x)}{K(x)}(\partial_x\theta)^2 \right], \qquad (3)$$

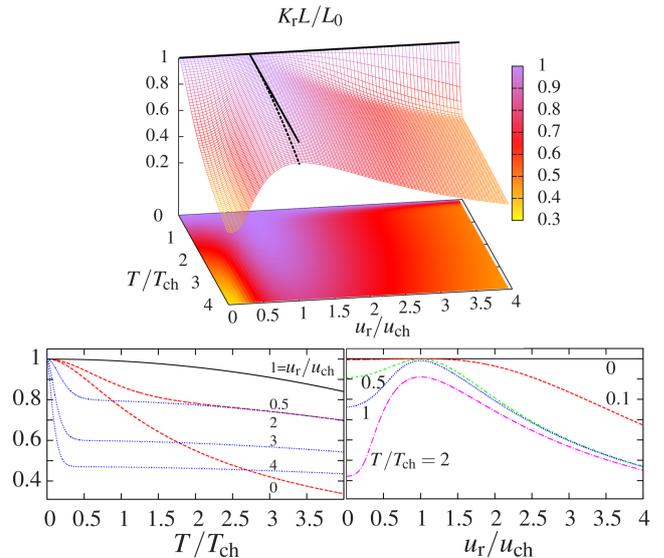

FIG. 2. Violation of the Wiedemann-Franz law. The Lorenz number (9) is plotted as a function of temperature and the ratio between the energy modes velocity in the reservoir and in the channel. Their momentum mismatch is quantified by the ratio $u_{\rm r}/u_{\rm ch}$. Energy modes are increasingly backscattered at increasing temperatures and for $u_{\rm r}/u_{\rm ch} \neq 1$, while particle transport is unaffected. This implies a suppression of the Lorenz number. Note that the WF law is violated also for $u_{\rm r} = u_{\rm ch}$ (compare solid with dashed black line) because of sub-leading interaction between energy modes yielding thermoelectric effects.

to describe the setup in Fig. 1. The velocity and interaction parameters $u(x)$ and $K(x)$ have different values $(u_{\rm ch}, K_{\rm ch})$ or $(u_{\rm r}, K_{\rm r})$, depending on whether $x$ is taken in the channel region $x \in \left[-\frac{d_{\rm ch}}{2}, \frac{d_{\rm ch}}{2}\right]$ or not. The LL parameters $u$ and $K$ change in the connection region between channel and reservoir. This connection is typically smooth on length scales of the order of the inter-particle distance $\sim 1/n_0$, but abrupt compared to the length wave of the low energy modes. This assumption allows to neglect particle backscattering in Eq. (3) and still consider a sharp variation of $u$ and $K$ which yields a scattering of energy modes. This different behavior of particle and energy transport leads to the violation of the WF law. In electronic systems, one typically deals with non interacting fermions in the reservoirs, implying $K_{\rm r} = 1$, and repulsive fermions in the reservoirs, i.e. $K_{\rm ch} < 1$. In the bosonic case, $K$ can have superunitary values both in the reservoirs and in the channel, with $K_{\rm r} \neq K_{\rm ch} \geq 1$, allowing to explore completely new parameter regimes.

We define the energy density $h(x)$ using Eq. (3) as $\mathcal{H}_0 = \int dx\, h(x)$. The continuity equations $\partial_x J + \partial_t n = 0$ and $\partial_x J_{\rm E} + \partial_t h = 0$ for the particle and energy densities

lead to the particle and energy current operators

$$J = \pi u K \Pi, \qquad J_E = -\frac{\hbar u^2}{2}\{\Pi, \partial_x \theta\}. \qquad (4)$$

The linear conductance is readily obtained as $g = K_r/h$; it is not renormalized by the interactions in the channel [21, 22, 25]. We use the Landauer-Büttiker theory of coherent transport [26] to derive the thermal conductivity $\mathcal{K}$. We diagonalize Eq. (3) with the help of bosonic scattering states of energy $\omega$ with transmission amplitude $t_\omega$ through the channel [21]. The energy current reads then $\langle J_E \rangle = \frac{\hbar}{2\pi} \int d\omega\, \omega |t_\omega|^2 [n_L(\omega) - n_R(\omega)]$, in which $n_\alpha(\omega)$ is the Bose distribution of the scattering modes in the reservoirs. The thermal conductivity $\mathcal{K}$ is defined in the $\Delta T \to 0$ limit of the energy current: $\langle J_E \rangle = \mathcal{K} \Delta T$ and the Lorenz number reads [11]

$$L_{LS} = \frac{\mathcal{K}}{gT} = \frac{L_0}{K_r} \frac{6}{\pi^2} \int_0^\infty dx |t_{2x/\hbar\beta}|^2 \frac{x^2}{\sinh^2(x)}. \qquad (5)$$

The label 'LS' stands for the linear spectrum assumption corresponding to Eq. (3). The transmission amplitudes $t_\omega$ depend only on frequencies $\omega$ and on the ratio $u_r/u_{ch}$, while they are insensitive to the Luttinger parameters $K_{r,ch}$. The reason is that backscattering is due to the breaking of translational invariance at the entrance of the channel and then to the momentum mismatch of the incoming wave $\omega/u$ [21]. For $u_r = u_{ch}$, no change of the eigenstates occurs along the cloud and $t_\omega = 1$, leading to $L_{LS} = L_0/K_r$. For $u_r \neq u_{ch}$, the transmission amplitudes $t_\omega$ become strongly energy dependent and control the deviation from the WF law at high temperatures. The Lorenz number $L_{LS}$ acquires a strong dependency on the velocity ratio $u_r/u_{ch}$ and on $T/T_{ch}$, in which $T_{ch} = \hbar u_r / 2 d_{ch} k_B$ is a characteristic temperature associated to the presence of the channel. Remarkably, a different behavior $L_{LS}$ is found for $u_r > u_{ch}$ (the case of electrons in metallic devices, see Fig. 1) and $u_r < u_{ch}$ (realizable with ultracold atoms). This is a manifestation of the wider range of possibilities offered by atomic setups. We note also that, in absence of chemical potential bias, the above approach yields a zero particle current for any temperature imbalance.

*Thermopower* - Thermoelectric effects are obtained once we go beyond the quadratic LL Hamiltonian (3); within the hydrodynamic approach, its corrections read [21]

$$\mathcal{H}_1 = -\int dx \left[ \frac{\pi \hbar^2}{8m} \{\Pi, \{\Pi, (\partial_x \theta)\}\} + \left.\frac{\delta^2 \mu}{\delta n^2}\right|_{n_0} \frac{(\partial_x \theta)^3}{6\pi^3} \right]. \qquad (6)$$

They describe an interaction between energy modes, and lead to a modified expression of the particle current operator

$$J = \pi u K \Pi + \frac{1}{mu^2} J_E. \qquad (7)$$

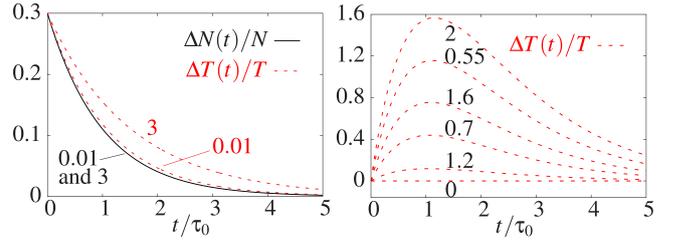

FIG. 3. Time evolution of the particle (solid black line) and temperature (dashed red lines) imbalances. Left: Imposing an initial particle or temperature imbalance, different time scales for particle and temperature equilibration are observed, as they are carried by different excitations. Temperature equilibration slows down for increasing temperatures: a manifestation of the suppression of the Lorenz number. The numbers indicate different ratios of $T/T_{ch}$. ($K_r = 1.18$, $u_r = 0.8 v_F$, $T/T_F = 0.2$ and $u_r/u_{ch} = 2.5$) Right: Tuning of thermoelectric effects by interactions in the channel. The presence of thermoelectric effects leads to an evolution of the temperature imbalance ($\Delta N(0)/N = 0.3$), given an initial particle imbalance. This is strongly sensitive to the momentum mismatch between collective modes at the entrance of the channel. This is controlled by the ratio $u_r/u_{ch}$.

To the same order of accuracy, the energy operator is unchanged. The second term in Eq. (7) couples mass and energy flows and is responsible for thermoelectric effects. It is controlled by the inverse mass, associated to the deviation of the spectrum from the linear dispersion. Using (7) we obtain that the thermopower is proportional to the thermal conductivity

$$s = \frac{1}{2}\left(\frac{v_F}{u_r}\right)^2 \frac{\mathcal{K}}{g E_F}, \qquad (8)$$

where $E_F = m v_F^2/2$ is the Fermi energy of the TG limit and $v_F = \pi \hbar n_0 / m$ is the Fermi velocity. An important consequence of $s$ being non zero is the modification of the Lorenz number (see Eq. (2)),

$$L = L_{LS}\left[1 - \frac{\pi^2}{12}\frac{L_{LS}}{L_0}\left(\frac{T}{T_F}\right)^2 \left(\frac{v_F}{u_r}\right)^4\right], \qquad (9)$$

with $T_F = E_F/k_B$. The dependence of the Lorenz number on temperature and the velocity ratio $u_r/u_{ch}$ is illustrated in Fig. 2. For non zero temperatures, the WF law is violated even in the homogeneous case $u_r = u_{ch}$, generalizing the known result for non-interacting fermions [4].

*Time scales* - Recent experiments [1] show the possibility to probe thermoelectric effects in the time evolution of the particle $\Delta N$ and temperature $\Delta T$ imbalances between the right and left reservoirs. Their time evolution is set by the differential equation

$$\tau_0 \partial_t \begin{pmatrix} \Delta N \\ \Delta T \end{pmatrix} = -\begin{pmatrix} 1 & -\kappa S \\ -\frac{S}{l\kappa} & \frac{L+S^2}{l} \end{pmatrix} \begin{pmatrix} \Delta N \\ \Delta T \end{pmatrix}. \qquad (10)$$

$\tau_0 = \kappa/g$ involves the compressibility of the reservoirs $\kappa = \partial N/\partial \mu|_T$, $l = C/\kappa T$ with specific heat $C = \partial E/\partial T|_N$, and $S = s_r - s$ the dilatation coefficient with $\kappa s_r = \partial N/\partial T|_\mu$. $\kappa$, $C$ and $s_r$ are the 'static' duals of the dynamic quantities $g$, $K$ and $s$ respectively. They can be extracted from the same correlation function by inverting the zero frequency limit $\omega \to 0$ and the zero momentum limit $q \to 0$ [21]. Assuming $d_r$ to be the size of the reservoirs, $\kappa = d_r K_r/\hbar\pi u_r$ and $C = k_B^2 d_r T\pi/3\hbar u_r$ [10].

We start by ignoring thermoelectric effects and setting $S = 0$. In this limit, $\Delta N(t) = e^{-\frac{t}{\tau_0}} \Delta N(0)$ while $\Delta T(t) = e^{-K_r \frac{L_{LS}}{L_0} \frac{t}{\tau_0}} \Delta T(0)$. Since in general $K_r L_{LS}/L_0 < 1$, we find that a temperature imbalance relaxes with a different time scale than a particle imbalance, see the left panel in Fig. 3. Taking now into account the corrections to the linear spectrum Eq. (6), we also derive [21] the dilatation coefficient $s_r = C/m u_r^2 \kappa$, leading to

$$S = \frac{L_0}{2k_B} \left(\frac{v_F}{u_r}\right)^2 \frac{T}{T_F} \frac{1}{K_r} \left[1 - \frac{K_r L_{LS}}{L_0}\right]. \qquad (11)$$

As shown in Fig. 3, the presence of thermoelectric effects could be probed by preparing a particle imbalance and measuring the time evolution of the temperature imbalance or *viceversa* and by changing the ratio $u_r/u_{ch}$. Remarkably, at low temperatures $T < T_F$, the particle exchange between reservoirs is much less sensitive to a initial temperature imbalance than the inverse case. This is due to the fact that the proportionality constant between $\Delta N(t)/N$ and an initial temperature imbalance $\Delta T(0)/T$ is proportional to $(T/T_F)^2$, which could be as low as $\sim 10^{-2}$, while, in the inverse case, the proportionality constant does not depend on temperature [21].

*Disorder effects-* In ultracold atomic setups, a tunable disorder can be added along the channel in a controllable way [1, 27]. We discuss here the consequences of disorder on bosonic transport. We identify two different regimes by comparing the typical disorder correlation length $l_D$ to the interparticle distance $n_0^{-1}$.

i) For $l_D \lesssim n_0^{-1}$, particle backscattering is unavoidable, and affects mass transport. A renormalization group (RG) analysis for delta-correlated disorder allows to determine its relevance at low energies in the presence of interactions [28]. Defining a dimensionless disorder strength $\mathcal{D}$ and a short distance cutoff $a$, the renormalization scaling transformation $a \to ae^l$ determines the flow of the disorder strength

$$\mathcal{D}(l) = \mathcal{D}e^{(3-2K_{ch})l}. \qquad (12)$$

For $K_{ch} > 3/2$ disorder is then irrelevant and the cloud remains in the superfluid phase. A weak disorder yields a renormalization of the LL parameter $K_{ch} \to K_{ch}^*$ and channel-length dependent corrections to the WF law [13]. In the high-temperature limit $k_B T \gg \hbar u_{ch}/d_{ch}$, the conductance reads [29]

$$g = \frac{K_r}{h}\left[1 - \mathcal{C}\, K_r \frac{d_{ch}}{a} \mathcal{D}\left(\frac{u_{ch}}{Ta}\right)^{2-2K_{ch}}\right], \qquad (13)$$

in which $\mathcal{C}$ is a non-universal factor depending on the UV regularization. The thermal conductivity $\mathcal{K}$ will depend non-universally on the details of the disorder. In the low temperature limit $T \ll u_{ch}/d_{ch}$ and exclusively in the TG limit for the reservoirs ($K_r = 1$) [29], the scaling of the $d_{ch}$-dependent corrections to the conductance can be derived by substituting $k_B T \Leftrightarrow \hbar u_{ch}/d_{ch}$ in Eq. (13) [13, 30].

Disorder becomes relevant in the RG sense for $K_{ch} < 3/2$. Below a localization temperature $T_{loc}$ the channel enters an insulating Bose glass phase [31]. The localization temperature tends to infinity in the TG limit $K_{ch} = 1$. For $T > T_{loc}$, the channel remains in the superfluid phase and perturbative results in the disorder strength $\mathcal{D}$ apply.

ii) For $l_D \gg n_0^{-1}$, particle backscattering can be neglected leaving mass transport unaffected, but long wavelength energy modes propagate in a random medium [20]. The disorder can be modeled by random fluctuations $\delta\rho$ of the static density $n_0(x) \to n_0(x) + \delta\rho(x)$, with $\langle\delta\rho(x)\delta\rho(x')\rangle = \langle\delta\rho(x)^2\rangle \exp(-(x-x')^2/l_D^2)$. Modes of energy $\omega$ acquire a mean free path [11, 20]

$$\xi(\omega) = \frac{4\, u_{ch}^6}{\sqrt{\pi}\, l_D \omega^2 A^2 \langle\delta\rho^2\rangle} e^{\omega^2 l_D^2/u_{ch}^2}. \qquad (14)$$

The factor $A$ is derived both in the strongly and weakly interacting limit by comparing the hydrodynamic approach to the exact solution of the Lieb-Liniger model [32]. Defining the dimensionless interaction strength $\gamma = mg/\hbar^2 n_0$, $A = 8n_0 (\pi\hbar/m)^2/\gamma$ for $\gamma \gg 1$ and $A = (\hbar/m)^2 n_0 \gamma$ for $\gamma \ll 1$ [21]. Note that the mean-free path diverges in the Tonks-Girardeau limit $\gamma \to \infty$ [33]: this result can be understood using the mapping onto a non-interacting Fermi gas, which is not scattered by a smooth disorder. Beyond the regime of validity of the Luttinger-Liquid picture, Eq. (14) should be matched at high energies with the free-particle behaviour $\xi(\omega) \propto \omega^2$ [33], leading to a non-monotonous dependence of the mean free path on the energy. A non-trivial regime of particle and energy transport may then occur when $\xi(\mu) > d_{ch}$: particles are not localized within the size of the sample, however, the energy modes may be localized if their energy $\omega_1$ is sufficiently large (though, still smaller than the chemical potential): in this case $\xi(\omega_1) < d_{ch}$. The energy modes do not conduct energy anymore, leading to a saturation of the thermal conductivity, or a suppression of the Lorenz number, for temperatures $T > \omega_1$. The energy scale $\omega_1$ has to be compared with the typical energy $\omega_2 \sim v_F/\sqrt{n_0^{-1} d_{ch}}$ at which the scattering states spontaneously decay because of the interaction term Eq.

(6). The condition $\omega_2/\omega_1 \gg 1$ sets the validity of the scattering approach presented in the previous discussion. We recall that we found $\langle J_{\rm E} \rangle \sim \Delta T$. If $T > \omega_2$, high energy modes are localized and interactions imply their decay into the low-energy ones, conducting heat. This down-energy conversion is responsible for an algebraic dependence of the energy current on the temperature imbalance [11, 34]

$$J_{\rm E} \sim \Delta T^{4/3} \quad \text{if} \quad \Delta T > \omega_2. \quad (15)$$

This power law is universal and it is an exquisite effect of interaction between energy modes.

To conclude, we showed the violation of the Wiedemann-Franz law and the presence of thermoelectric effects in 1D cold-atom clouds of strongly interacting bosons. Particle-energy separation is signaled by different time evolution of particle or temperature imbalances between two 1D reservoirs connected by a 1D channel. Our results generalize and extend previous results valid for fermions in 1D solid state devices. Strongly interacting bosons in one-dimension open new scenarios to explore transport in low-dimensional nano-structures and probe various quantum phases, including many-body localization [31].

We thank D. Basko for discussions and M. Schneider for useful comments on the paper. We acknowledge financial support from the Alexander von Humboldt Foundation, the Institut Universitaire de France, the ANR project no. ANR-13-JS01-0005-01 and the ERC Handy-Q grant no. 25860.

# Supplemental Material for "Violation of the Wiedemann-Franz Law for ultracold atomic gases"

Michele Filippone[1], Frank Hekking[2], and Anna Minguzzi[2]

[1] Dahlem Center for Complex Quantum Systems and Institut für Theoretische Physik,
Freie Universität Berlin, Arnimallee 14, 14195 Berlin, Germany and
[2] Université Grenoble 1/CNRS, Laboratoire de Physique et de Modélisation
des Milieux Condensés (UMR 5493), B.P. 166, 38042 Grenoble, France

In this Supplemental Material, we provide the detailed discussion of the techniques developed and the calculations carried out to derive the analytical results in the main text.

## S-I. MODEL

The corrections to the linearized effective model are derived to the next to leading order in the hydrodynamic approach of Refs. [1]. This consist in deriving the effective Hamiltonian of a quantum 1D fluid from the classical equations of hydrodynamics

$$\partial_t n + \partial_x(vn) = 0, \qquad mn d_t v = -n\partial_x \mu[n] - n\partial_x V_{\text{ext}}; \qquad \text{(S-1)}$$

in which $n(x)$ and $v(x)$ are the fluid density and velocity respectively and $\mu[n]$ is a potential functional of $n$, connected to the fluid pressure via $\partial_x P = n\partial_x \mu[n]$. The first is the continuity equation and the second Newton's equation. The linearization close to equilibrium is performed by setting $n(x,t) = n_0(x) + \delta n(x,t)$ and $v(x,t) = \partial_t \vartheta(x,t)$, in which $\delta n$ and $\vartheta$ are both next to leading order fields and $\vartheta$ is a displacement field. Equating order by order, the continuity equation leads to

$$\delta n = -\partial_x(n_0 \vartheta), \qquad \partial_x[\delta n \partial_t \vartheta] = 0. \qquad \text{(S-2)}$$

The functional $\mu[n]$ is expanded close to the equilibrium value $n_0$: $\mu[n] = \mu[n_0] + \delta n (\delta\mu/\delta n)|_{n_0} + \delta n^2 (\delta^2\mu/\delta n^2)|_{n_0}/2\ldots$ To zeroth order, the thermal equilibrium condition is derived

$$\partial_x\Big(\mu[n_0] + V_{\text{ext}}\Big) = 0, \qquad \Rightarrow \qquad \mu[n_0] + V_{\text{ext}} = \mu, \qquad \text{(S-3)}$$

in which $\mu$ is the physical chemical potential. Making explicit the Lagrangian derivative $d_t = \partial_t + v\partial_x$, to first order, Newton's equation reads

$$mn_0 \partial_t^2 \vartheta + n_0 \partial_x \left[\partial_x(n_0 \vartheta) \left.\frac{\delta\mu}{\delta n}\right|_{n_0}\right] = 0, \qquad \text{(S-4)}$$

in which Eq. (S-2) has been applied. This is multiplied by $\partial_t \vartheta$, integrated by $x$ and a time-conserved quantity is derived, which is identified to the energy of the system

$$\varepsilon = \int dx \left\{\frac{mn_0}{2}(\partial_t \vartheta)^2 + \frac{1}{2}\left.\frac{\delta\mu}{\delta n}\right|_{n_0} [\partial_x(n_0 \vartheta)]^2\right\}. \qquad \text{(S-5)}$$

The momentum associated to $\vartheta$ reads $p(x,t) = mn_0 \partial_t \vartheta$. The quantization of this Hamiltonian then requires $[u(x), p(x')] = i\hbar \delta(x-x')$. Adopting the convention $\delta n = -\partial \theta/\pi$, one can write this Hamiltonian in the standard bosonization notation

$$\mathcal{H}_0 = \frac{\hbar}{2\pi}\int dx \left\{u(x)K(x)\pi^2 \Pi(x)^2 + \frac{u(x)}{K(x)}[\partial_x \theta(x)]^2\right\}, \qquad \text{(S-6)}$$

with

$$\theta(x) = \pi n_0(x)\vartheta(x), \qquad \Pi(x) = \frac{m}{\pi\hbar}\partial_t \vartheta(x), \qquad u(x)K(x) = \frac{\hbar\pi}{m}n_0(x), \qquad \frac{u(x)}{K(x)} = \frac{1}{\pi\hbar}\left.\frac{\delta\mu}{\delta n}\right|_{n_0} \qquad \text{(S-7)}$$

and $[\theta(x), \Pi(x')] = i\delta(x-x')$. The next to leading order contribution to Newton's equation reads

$$mn_0 \partial_t \vartheta \partial_x \partial_t \vartheta + m\delta n \partial_t^2 \vartheta + n_0 \partial_x\left[\frac{\delta n^2}{2}\frac{\delta^2 \mu}{\delta n^2}\right] + \delta n \partial_x\left[\delta n \frac{\delta \mu}{\delta n}\right] = 0. \qquad \text{(S-8)}$$

The second and fourth term simplify as they coincide with Eq. (S-4), leading to

$$mn_0\partial_t\vartheta\partial_x\partial_t\vartheta + n_0\partial_x\left[\frac{\delta n^2}{2}\frac{\delta^2\mu}{\delta n^2}\right] = 0\,. \tag{S-9}$$

This equation is multiplied again by $\partial_t\vartheta$ and integrated by $x$. The energy correction is obtained after the following set of manipulations (remember $\delta n = -\partial_x(n_0\vartheta)$)

$$\begin{aligned}
\int dx \partial_t(n_0\vartheta)\partial_x\left[\frac{\delta n^2}{2}\frac{\delta^2\mu}{\delta n}\right] &= -\int dx[\partial_t\partial_x(n_0\vartheta)]\frac{(\partial_x n_0\vartheta)^2}{2}\frac{\delta^2\mu}{\delta n^2} = -\partial_t\int dx \frac{1}{6}\frac{\delta^2\mu}{\delta n^2}[\partial_x(n_0\vartheta)]^3\,,\\
\int dx m n_0(\partial_t\vartheta)^2\partial_x\partial_t\vartheta &= -\int dx m(\partial_t\partial_x n_0\vartheta)\frac{(\partial_t\vartheta)^2}{2}\\
&= -\partial_t\int dx m(\partial_x n_0\vartheta)\frac{(\partial_t\vartheta)^2}{2} + \int dx m\partial_x(n_0\vartheta)\partial_t\vartheta\partial_t^2\vartheta\,.
\end{aligned} \tag{S-10}$$

Eq. (S-2) implies that $\partial_x(n_0\vartheta)\partial_t\vartheta$ does not depend on $x$ and $\partial_t^2\vartheta$ is a global derivative of $x$ because of Eq. (S-4). The last term in the last equality is then zero and does not contribute. The correction to the time conserved quantity Eq. (S-5) reads then

$$\varepsilon_1 = -\int dx\left\{m\partial_x(n_0\vartheta)\frac{(\partial_t\vartheta)^2}{2} + \frac{1}{6}\frac{\delta^2\mu}{\delta n^2}[\partial_x(n_0\vartheta)]^3\right\}\,. \tag{S-11}$$

Applying the set of transformations Eq. (S-7) and after symmetrization, the correction to the Hamiltonian Eq. (3) reads

$$\mathcal{H}_1 = -\int dx\left\{\frac{\pi\hbar^2}{8m}\{\Pi(x),\{\Pi(x),\partial_x\theta(x)\}\} + \frac{1}{6\pi^3}\left.\frac{\delta^2\mu}{\delta n^2}\right|_{n=n_0}[\partial_x\theta(x)]^3\right\}\,. \tag{S-12}$$

To leading order, the energy current is obtained from the continuity equation $\partial_x J_\text{E} + \partial_t \langle(x) = 0$, $h(x)$ being the energy density, such that $\mathcal{H}_0 = \int dx h(x)$. The energy current reads

$$J_\text{E} = -\hbar\frac{u(x)^2}{2}\{\Pi(x),\partial_x\theta(x)\}\,. \tag{S-13}$$

The particle current is derived in the same way from the continuity equation for the particle density $\partial_x J + \partial_t n = 0$. Also taking into account the next to leading order correction Eq. (6) are taken into account. The continuity equation implies $J = \partial_t\theta/\pi$, such that

$$J = u(x)K(x)\Pi(x) - \frac{\hbar}{2m}\{\Pi(x),\partial_x\theta(x)\} = u(x)K(x)\Pi(x) + \frac{1}{mu^2(x)}J_\text{E}\,, \tag{S-14}$$

which recovers, with Eq. (S-13), Eqs. (4) and (7) in the main text.

### S-II.  TRANSPORT COEFFICIENTS

We determine now the coefficients of the transport matrix appearing in Eq. (2) in the main text.

#### A.  Conductance

The linear conductance of the system is found extending the steps of Refs. [2] to the case in which $K > 1$. In the Matsubara formalism, in which one performs a transformation to imaginary time $it = \tau$, the conductance is given for any temperature by the Kubo formula ($\hbar = 1$) [3]

$$g_{xx'} = -\lim_{\nu_n\to 0}\frac{1}{\nu_n}\int_0^\beta d\tau e^{i\nu_n\tau}\langle T_\tau J(x,\tau)J(x',0)\rangle\,, \tag{S-15}$$

in which $x'$ is any point within the channel, along which the chemical potential switch from the left to the right value ($\mu_L$ to $\mu_R$). The continuity equation sets $J(i\nu_n) = \nu_n \theta(i\nu_n)/\pi$, allowing to write the conductance as the zero frequency limit of the correlation function of the density-fluctuations

$$g_{xx'} = \lim_{i\nu_n \to i0^+} \frac{\nu_n}{\pi^2} G_{\theta\theta}(x, x', i\nu_n), \qquad G_{\theta\theta}(x, x', i\nu_n) = \langle \theta(x, i\nu_n)\theta(x', -i\nu_n)\rangle. \tag{S-16}$$

The equation of motion for $G_{\theta\theta}$ reads

$$\left[-\partial_x \frac{u}{K}\partial_x + \frac{\nu_n^2}{uK}\right] G_{\theta\theta}(x, x', i\nu_n) = \pi\delta(x - x'), \tag{S-17}$$

implying that the first derivative of $G_{\theta\theta}$ is discontinuous at $x = x'$

$$\pi = -\left.\frac{u(x)}{K(x)}\partial_x G_{\theta\theta}(i\nu_n, x, x')\right|_{x'-0^+}^{x=x'+0^+}. \tag{S-18}$$

Setting open boundary conditions $G(x = \pm\infty, i\nu_n) = 0$, these equations are satisfied by the ansatz

$$G_{\theta\theta}(i\nu_n, x, x') = \begin{cases} A(i\nu_n, x')e^{\frac{|\nu_n|x}{u_{\rm r}}} & x \leq \frac{d_{\rm ch}}{2} \\ B(i\nu_n, x')e^{\frac{|\nu_n|x}{u_{\rm ch}}} + C(i\nu_n, x')e^{-\frac{|\nu_n|x}{u_{\rm ch}}} & -\frac{d_{\rm ch}}{2} < x \leq x' \\ D(i\nu_n, x')e^{\frac{|\nu_n|x}{u_{\rm ch}}} + E(i\nu_n, x')e^{-\frac{|\nu_n|x}{u_{\rm ch}}} & x' < x \leq \frac{d_{\rm ch}}{2} \\ F(i\nu_n, x')e^{-\frac{|\nu_n|x}{u_{\rm r}}} & \frac{d_{\rm ch}}{2} < x, \end{cases} \tag{S-19}$$

The solution is given by

$$\begin{aligned}
A(|\nu_n|, x') &= \frac{\pi}{|\nu_n|} \frac{K_{\rm ch} K_{\rm r} e^{\frac{d_{\rm ch}|\nu_n|}{2u_{\rm r}}}}{K_{\rm ch} f(d_{\rm ch}) + K_{\rm r} g(d_{\rm ch})} f\left(\frac{d_{\rm ch}}{2} - x'\right) \\
B(|\nu_n|, x') &= \frac{\pi}{2|\nu_n|} \frac{K_{\rm ch}(K_{\rm ch} + K_{\rm r}) e^{\frac{d_{\rm ch}|\nu_n|}{2u_{\rm r}}}}{K_{\rm ch} f(d_{\rm ch}) + K_{\rm r} g(d_{\rm ch})} f\left(\frac{d_{\rm ch}}{2} - x'\right) \\
C(|\nu_n|, x') &= \frac{\pi}{2|\nu_n|} \frac{K_{\rm ch}(K_{\rm r} - K_{\rm ch}) e^{-\frac{d_{\rm ch}|\nu_n|}{2u_{\rm r}}}}{K_{\rm ch} f(d_{\rm ch}) + K_{\rm r} g(d_{\rm ch})} f\left(\frac{d_{\rm ch}}{2} - x'\right) \\
D(|\nu_n|, x') &= \frac{\pi}{2|\nu_n|} \frac{K_{\rm ch}(K_{\rm r} - K_{\rm ch}) e^{-\frac{|\nu_n|d_{\rm ch}}{2u_{\rm ch}}}}{K_{\rm ch} f(d_{\rm ch}) + K_{\rm r} g(d_{\rm ch})} f\left(x' + \frac{d_{\rm ch}}{2}\right) \\
E(|\nu_n|, x') &= \frac{\pi}{2|\nu_n|} \frac{K_W(K_L + K_W) e^{\frac{|\nu_n|d_{\rm ch}}{u_{\rm ch}}}}{K_{\rm ch} f(d_{\rm ch}) + K_{\rm r} g(d_{\rm ch})} f\left(x' + \frac{d_{\rm ch}}{2}\right) \\
F(|\nu_n|, x') &= \frac{\pi}{|\nu_n|} \frac{K_W K_L e^{\frac{|\nu_n|d_{\rm ch}}{u_{\rm r}}}}{K_{\rm ch} f(d_{\rm ch}) + K_{\rm r} g(d_{\rm ch})} f\left(x' + \frac{d_{\rm ch}}{2}\right),
\end{aligned} \tag{S-20}$$

in which we defined the functions

$$f(x) = (K_{\rm r} + K_{\rm ch})e^{\frac{x|\nu_n|}{u_{\rm ch}}} + (K_{\rm r} - K_{\rm ch})e^{\frac{-x|\nu_n|}{u_{\rm ch}}}, \qquad g(x) = (K_{\rm r} + K_{\rm ch})e^{\frac{x|\nu_n|}{u_{\rm ch}}} - (K_{\rm r} - K_{\rm ch})e^{\frac{-x|\nu_n|}{u_{\rm ch}}}. \tag{S-21}$$

To obtain the linear conductance in Eq. (S-16), we take the limit $i\nu_n \to i0^+$. Noticing that $G_{\theta\theta} \to \pi K_{\rm r}/\nu_n$, the conductance in Eq. (S-16) does not depend on $x$ and $x'$ and it reads $g = K_{\rm r}/2\pi = K_{\rm r}/h$ (the last equality reestablishes physical dimensions). This result is insensitive to interactions in the channel.

### B. Thermal conductivity

Eq. (3) is diagonalized by making the transformation

$$\begin{aligned}
\theta(x) &= \sum_{\alpha=L,R} \int_0^\infty d\omega \sqrt{\frac{\pi u(x) K(x)}{2\omega}} \left(\phi_{\omega\alpha}(x) a_{\omega\alpha} + \phi^*_{\omega\alpha}(x) a^\dagger_{\omega\alpha}\right), \\
\Pi(x) &= \sum_{\alpha=L,R} \int_0^\infty d\omega \sqrt{\frac{\omega}{2\pi u(x) K(x)}} \frac{\phi_{\omega\alpha}(x) a_{\omega\alpha} - \phi^*_{\omega\alpha}(x) a^\dagger_{\omega\alpha}}{i},
\end{aligned} \tag{S-22}$$

in which the bosonic operators $a^\dagger_{\omega\alpha}$ satisfy to $\left[a_{\omega\alpha}, a^\dagger_{\omega'\alpha'}\right] = \delta_{\omega\omega'}\delta_{\alpha\alpha'}$ and create excitations in the scattering eigenstates $\phi_{\omega\alpha}$ of energy $\hbar\omega$, coming from the left/right reservoir ($\alpha = L/R$). The eigenfunctions $\phi_{\omega\alpha}$ satisfy to

$$h_0 \phi_{\omega\alpha}(x) = -\sqrt{u(x)K(x)}\partial_x \frac{u(x)}{K(x)} \partial_x \sqrt{u(x)K(x)} \, \phi_{\omega\alpha}(x) = \omega^2 \phi_{\omega\alpha}(x), \tag{S-23}$$

leading to the diagonal form for Eq. (3) in the main text

$$\mathcal{H}_0 = \sum_{\alpha=L,R} \int_0^\infty d\omega \, \hbar\omega \left(a^\dagger_{\omega\alpha} a_{\omega\alpha} + \frac{1}{2}\right). \tag{S-24}$$

The scattering basis, solution of Eq. (S-23), can be written into the form

$$\phi_{\omega L}(x) = \frac{1}{\sqrt{2\pi u_r}} \begin{cases} e^{i\frac{\omega}{u_r}x} + r_\omega e^{-i\frac{\omega}{u_r}x} & x < -\frac{d_{\rm ch}}{2} \\ a_\omega e^{i\frac{\omega}{u_{\rm ch}}x} + b_\omega e^{-i\frac{\omega}{u_{\rm ch}}x} & -\frac{d_{\rm ch}}{2} < x < \frac{d_{\rm ch}}{2} \\ t_\omega e^{i\frac{\omega}{u_r}x} & \frac{d_{\rm ch}}{2} < x \end{cases},$$

$$\phi_{\omega R}(x) = \frac{1}{\sqrt{2\pi u_r}} \begin{cases} t_\omega e^{-i\frac{\omega}{u_r}x} & x < -\frac{d_{\rm ch}}{2} \\ c_\omega e^{i\frac{\omega}{u_{\rm ch}}x} + d_\omega e^{-i\frac{\omega}{u_{\rm ch}}x} & -\frac{d_{\rm ch}}{2} < x < \frac{d_{\rm ch}}{2} \\ e^{i\frac{\omega}{u_r}x} - \frac{r_\omega^* t_\omega}{t_\omega^*} e^{-i\frac{\omega}{u_r}x} & \frac{d_{\rm ch}}{2} < x \end{cases}. \tag{S-25}$$

Energy is conserved all along the system. The energy current can be then calculated at any point. Choosing $x = \to -\infty$, we obtain

$$\langle J_{\rm E}(x=-\infty)\rangle = -\hbar \frac{u_r^2}{2} \sum_\alpha \int_0^\infty dk \, (2n_{k\alpha}+1) \, {\rm Im}\left[\phi_{k\alpha}(x)\partial_x \phi^*_{k\alpha}(x)\right] = \frac{\hbar}{2\pi} \int_0^\infty d\omega \, \omega \left[n_L(\omega) - n_R(\omega)\right] |t_\omega|^2, \tag{S-26}$$

leading to Eq. (5) in the main text for the Lorenz number. In the following, we provide an explicit expression for the transmission amplitudes $t_\omega$ in the case of a sharp transition of the Luttinger parameters, such that $(u,K) = (u_{\rm ch}, K_{\rm ch})$ if $x \in [-d_{\rm ch}/2, d_{\rm ch}/2]$ and $(u,K) = (u_r, K_r)$ otherwise. The eigenvalue equation Eq. (S-23) implies the continuity of the function $\Psi_{\omega\alpha}(x) = \sqrt{u(x)K(x)}\phi_{\omega\alpha}(x)$ with a discontinuous derivative, satisfying to the following boundary condition at the junction between the reservoir and the channel

$$u_r^2 \partial_x \Psi_r = u_{\rm ch}^2 \partial_x \Psi_{\rm ch}. \tag{S-27}$$

These conditions are satisfied by a solution of the form

$$\Psi_{\omega L}(x) = \frac{\sqrt{u_r K_r}}{\sqrt{2\pi u_r}} \begin{cases} e^{i\frac{\omega}{u_r}x} + r_\omega e^{-i\frac{\omega}{u_L}x} & x < -\frac{d_{\rm ch}}{2} \\ \sqrt{\frac{u_{\rm ch}K_{\rm ch}}{u_r K_r}} \left[a_\omega e^{i\frac{\omega}{u_{\rm ch}}x} + b_\omega e^{-i\frac{\omega}{u_{\rm ch}}x}\right] & -\frac{d_{\rm ch}}{2} < x < \frac{d_{\rm ch}}{2} \\ t_\omega e^{i\frac{\omega}{u_r}x} & \frac{d_{\rm ch}}{2} < x \end{cases},$$

$$\Psi_{\omega R}(x) = \frac{\sqrt{u_r K_r}}{\sqrt{2\pi u_r}} \begin{cases} t_\omega e^{-i\frac{\omega}{u_r}x} & x < -\frac{d_{\rm ch}}{2} \\ \sqrt{\frac{u_{\rm ch}K_{\rm ch}}{u_r K_r}} \left[c_\omega e^{i\frac{\omega}{u_{\rm ch}}x} + d_\omega e^{-i\frac{\omega}{u_{\rm ch}}x}\right] & -\frac{d_{\rm ch}}{2} < x < \frac{d_{\rm ch}}{2} \\ e^{i\frac{\omega}{u_r}x} - \frac{r_\omega^* t_\omega}{t_\omega^*} e^{-i\frac{\omega}{u_r}x} & \frac{d_{\rm ch}}{2} < x \end{cases}. \tag{S-28}$$

The continuity of $\Psi_{\omega\alpha}(x)$ plus the matching condition for the derivative Eq. (S-27) lead to

$$r_\omega = \frac{(u_r - u_{\rm ch})(u_r + u_{\rm ch})e^{-\frac{id_{\rm ch}\omega}{u_r}}\left(-1 + e^{\frac{2id_{\rm ch}\omega}{u_{\rm ch}}}\right)}{-(u_r + u_{\rm ch})^2 + (u_r - u_{\rm ch})^2 e^{\frac{2id_{\rm ch}\omega}{u_{\rm ch}}}},$$

$$a_\omega = -\frac{2u_r(u_r + u_{\rm ch})e^{\frac{id_{\rm ch}\omega(u_r - u_{\rm ch})}{2u_r u_{\rm ch}}}}{-(u_r + u_{\rm ch})^2 + (u_r - u_{\rm ch})^2 e^{\frac{2id_{\rm ch}\omega}{u_{\rm ch}}}} \sqrt{\frac{u_r K_r}{u_{\rm ch} K_{\rm ch}}},$$

$$b_\omega = \frac{2u_r(u_r - u_{\rm ch})e^{\frac{id_{\rm ch}\omega(3u_r - u_{\rm ch})}{2u_r u_{\rm ch}}}}{-(u_r + u_{\rm ch})^2 + (u_r - u_{\rm ch})^2 e^{\frac{2id_{\rm ch}\omega}{u_{\rm ch}}}} \sqrt{\frac{u_r K_r}{u_{\rm ch} K_{\rm ch}}},$$

$$t_\omega = -\frac{4u_r u_{\rm ch} e^{\frac{id_{\rm ch}\omega(u_r - u_r)}{u_r u_{\rm ch}}}}{-(u_r + u_{\rm ch})^2 + (u_r - u_{\rm ch})^2 e^{\frac{2id_{\rm ch}\omega}{u_r}}}. \tag{S-29}$$

We focus on the transmission coefficients as they appear in Eq. (5) in the main text, that is

$$t_{2x/\hbar\beta} = -\frac{4\nu e^{iT^*x(\nu-1)}}{-(\nu+1)^2 + (\nu-1)^2 e^{2iT^*x\nu}}, \qquad \nu = \frac{u_r}{u_{ch}}, \qquad T^* = \frac{T}{T_{ch}}, \qquad T_{ch} = \frac{\hbar u_r}{2d_{ch}k_B}. \tag{S-30}$$

This formulation of the transmission amplitudes stresses that the presence of the channel is responsible for the emergence of a further energy scale $k_B T_{ch}$ and that the transmission amplitudes are insensitive to the mismatch of the interaction parameters $K$ of the Luttinger liquid between channel and reservoirs, but only the bosonic velocities $u$. The reason is that backscattering of free waves is provoked by the breaking of translational symmetry at the channel. It is then exclusively controlled by the momentum mismatch $\omega/u_r$ to $\omega/u_{ch}$ of the the plane waves from the reservoirs to the channel, which is insensitive to the interaction parameter $K$.

## S-III. THERMODYNAMIC COEFFICIENTS FROM CORRELATION FUNCTIONS

The aim of this section is to show how it is possible to derive the dilatation coefficient of a Luttinger liquid from the correlation functions making use of the correction to the particle current appearing in Eq. (7) in the main text, a manifestation of the correction to the linear spectrum approximation derived in Eq. (6). We first derive, for pedagogical purpose, the compressibility $\kappa$ and the specific heat $C$ from the (density)-(density) and (energy density)-(energy density) correlation function respectively. We show then how these results can be used to derive the dilatation coefficient $s_r$.

### A. Compressibility

The compressibility

$$\kappa = \left.\frac{\partial N}{\partial \mu}\right|_T = \beta\left[\langle N^2\rangle - \langle N\rangle^2\right] \tag{S-31}$$

can be obtained as the zero frequency limit of the response function

$$K_{NN}(i\nu_n) = \int_0^\beta d\tau e^{i\nu_n\tau} K_{NN}(\tau), \qquad K_{NN}(\tau-\tau') = \left[\langle T_\tau N(\tau)N(\tau')\rangle - \langle N\rangle^2\right], \tag{S-32}$$

in which $i\nu_n = 2\pi n/\beta$ ($n$ integer) are bosonic Matsubara frequencies ensuring the periodicity of bosonic fields in imaginary time $t = i\tau$ and $T_\tau$ is the imaginary time ordering operator. Making the Fourier transform, the time independent averages of $\langle N\rangle^2$ is zero, leading to

$$K_{NN}(i\nu_n) = \langle N(i\nu_n)N(-i\nu_n)\rangle = d_r \lim_{k\to 0}\langle n(k,i\nu_n)n(-k,-i\nu_n)\rangle = d_r \lim_{k\to 0}\frac{k^2}{\pi^2}G_{\theta\theta}(k,i\nu_n). \tag{S-33}$$

In the last equality we exploit the fact that $n = -\partial_x\theta/\pi$, leading to the correlation function defined in Eq. (S-16). In the uniform case, the solution of the equation of motion Eq. (S-17) is readily found

$$G_{\theta\theta}(k,i\nu_n) = \frac{\pi}{\frac{u}{K}k^2 + \frac{\nu_n^2}{uK}}, \tag{S-34}$$

leading to

$$K_{NN}(i\nu_n) = d_r \lim_{k\to 0}\frac{k^2}{\pi^2}\frac{\pi}{\frac{u}{K}k^2 + \frac{\nu_n^2}{uK}}. \tag{S-35}$$

Taking the zero frequency limit (static chemical potential to get pa thermodynamic response ) *before* the $k\to 0$ limit (uniform potential) one finds, reestablishing dimensions, the known result for the compressibility [4]

$$\kappa = \frac{d_r}{\hbar}\frac{K_r}{\pi u_r}. \tag{S-36}$$

## B. Specific heat

The same kind of arguments applies for the specific heat:

$$C = \left.\frac{\partial \langle \mathcal{H} \rangle}{\partial T}\right|_N = k_B \beta^2 \left[\langle \mathcal{H}^2 \rangle - \langle \mathcal{H} \rangle^2\right]. \tag{S-37}$$

In analogy with the compressibility derived in Sec. S-III A, one has to study the zero frequency and momentum limit of the correlation function

$$K_{HH}(k, i\nu_n) = \langle h(k, i\nu_n) h(-k, -i\nu_n) \rangle, \tag{S-38}$$

in which

$$h(x) = \frac{1}{2\pi}\left[uK(\partial_x\phi)^2 + \frac{u}{K}(\partial_x\theta)^2\right] \tag{S-39}$$

is the energy density giving the total Hamiltonian $\mathcal{H}_0 = \int dx\, h(x)$, Eq. (3) in the main text. We stressed the presence of the field $\phi(x)$, defined as such $\partial_x \phi = \pi \Pi$, which is dual to the density-fluctuation field $\theta$ and for which the correlation function can be readily obtained by multiplying Eq. (S-34) by $1/K^2$

$$G_{\phi\phi}(k, i\nu_n) = \frac{u}{K}\frac{\pi}{u^2 k^2 + \nu_n^2}. \tag{S-40}$$

In the reciprocal space the energy density reads

$$h(k, i\nu_n) = \frac{u}{2\pi} \sum_{p, i\omega_n} p(p-k) \left[K \phi(p, i\omega_n)\phi(k-p, i\nu_n - i\omega_n) + \frac{\theta(p, i\omega_n)\theta(k-p, i\nu_n - i\omega_n)}{K}\right], \tag{S-41}$$

and the specific heat is then by contributions of the form

$$\frac{C}{4} = \beta \left(\frac{u}{2\pi}\right)^2 \sum_{p, p', i\omega_n, i\omega'_n} p(p-k)p'(p'+k)K^2 \langle \phi(p, i\omega_n)\phi(k-p, i\nu_n - i\omega_n)\phi(p', i\omega'_n)\phi(-k-p', -i\nu_n - i\omega'_n) \rangle. \tag{S-42}$$

The factor 4 in the left hand side is put to stress that the 3 further contributions involving averages like $\langle \phi\phi\theta\theta \rangle$, $\langle \theta\theta\phi\phi \rangle$ and $\langle \theta\theta\theta\theta \rangle$ lead exactly to the same result [5]. We apply Wick's theorem. The contraction $\overline{\phi\phi}\overline{\phi\phi}$ is zero because $i\nu_n \neq 0$ (this limit must me carried out at the very end of calculations) and one has only to consider $\phi\phi\phi\phi$ and $\phi\phi\phi\phi$. Both give the same contribution

$$\frac{C}{4} = \left(\frac{u}{2\pi}\right)^2 \sum_{p, i\omega_n} p^2(p-k)^2 K^2 \langle \phi(p, i\omega_n)\phi(-p, -i\omega_n) \rangle \langle \phi(k-p, i\nu_n - i\omega_n)\phi(-k+p, -i\nu_n + i\omega'_n) \rangle. \tag{S-43}$$

Inserting Eq. (S-40) the Matsubara sum can be carried out leading to

$$\frac{C}{4} = -\beta \frac{u^4}{4} \sum_p p^2(p-k)^2 \left[\frac{n(up)}{2up(uk - i\nu_n)(2up - uk - i\nu_n)} - \frac{n(-up)}{2up(uk - 2up - i\nu_n)}\right.$$
$$\left. - \frac{n(up - uk)}{uk(2up - uk)(2up - 2uk - i\nu_n)} + \frac{n(-up + uk)}{uk(uk - up)(2uk - 2up - i\nu_n)}\right], \tag{S-44}$$

in which $n(up) = 1/(e^{\beta up} - 1)$ is the Bose distribution. The limit $i\nu_n \to 0$ and $k \to 0$ can be subsequently carried out leading to (reestablishing dimensions)

$$\frac{C}{4} = \frac{k_B \beta}{\hbar}\frac{u^4}{2}\sum_{p>0}\frac{\beta p^2}{4u^2 \sinh^2(\beta up/2)} = \frac{k_B d_r}{\hbar \beta}\frac{1}{2\pi}\frac{1}{u}\int_0^\infty dx \frac{x^2}{\sinh^2(x)}. \tag{S-45}$$

Given that $\int_0^\infty dx \frac{x^2}{\sinh^2(x)} = \pi^2/6$, we find the known result [4]

$$C = \frac{k_B^2 T d_r}{\hbar u_r}\frac{\pi}{3}. \tag{S-46}$$

### C. Dilatation coefficient

The dilatation coefficient $s_\mathrm{r}$ is defined through

$$\kappa\, s_\mathrm{r} = \left.\frac{\partial N}{\partial T}\right|_\mu = k_B \beta^2 \left[\langle NH \rangle - \langle N \rangle \langle H \rangle\right],\tag{S-47}$$

and requires then the study of the correlation function

$$K_{NH}(k, i\nu_n) = \langle n(k, i\nu_n) h(-k, -i\nu_n) \rangle\,.\tag{S-48}$$

This is normally zero if only the particle-hole symmetric Hamiltonian Eq. (3) is taken into account. To consider the sub-leading contribution provided by the correction to the particle current operator in Eq. (7) in the main text, we rely on the continuity equation to show that

$$n(k, i\nu_n) = -\frac{ik}{\nu_n}\left[J + \frac{1}{2}\frac{v_F^2}{u^2}\frac{1}{E_F}J_\mathrm{E}\right] = -\frac{ik}{\nu_n}J + \frac{1}{2}\frac{v_F^2}{u^2}\frac{1}{E_F}h(k, i\nu_n)\,.\tag{S-49}$$

This automatically leads to

$$K_{NH} = \frac{1}{2}\frac{v_F^2}{u^2}\frac{1}{E_F}K_{HH}\tag{S-50}$$

and the dilatation coefficient

$$s_r = \frac{1}{2}\frac{v_F^2}{u_\mathrm{r}^2}\frac{1}{E_F}\frac{C}{\kappa} = \frac{T\pi^2 k_B^2}{3m K_\mathrm{r} u_\mathrm{r}^2}\,.\tag{S-51}$$

This allows to obtain Eq. (11) in the main text,

$$S = \frac{L_0}{2k_B}\left(\frac{v_F}{u_\mathrm{r}}\right)^2 \frac{T}{T_F}\frac{1}{K_\mathrm{r}}\left[1 - \frac{K_\mathrm{r} L_\mathrm{LS}}{L_0}\right]\,.\tag{S-52}$$

We note that $S \geq 0$ and becomes zero if energy mode backscattering is suppressed, that is for $u_\mathrm{r} = u_\mathrm{ch}$.

## S-IV. TIME EVOLUTION OF TEMPERATURE AND PARTICLE IMBALANCES

The solution of Eq. (10) in the main text is given by

$$\begin{aligned}\Delta N(\tau) =& \frac{1}{2}\left\{\left(e^{-\lambda_+ \tau} + e^{-\lambda_- \tau}\right) - \frac{2 - \lambda_+ - \lambda_-}{(\lambda_+ - \lambda_-)}\left(e^{-\lambda_- \tau} - e^{-\lambda_+ \tau}\right)\right\}\Delta N(0) + \\ & + \frac{\kappa S}{\lambda_+ - \lambda_-}\left(e^{-\lambda_- \tau} - e^{-\lambda_+ \tau}\right)\Delta T(0)\,,\\ \Delta T(\tau) =& \frac{1}{2}\left\{\left(e^{-\lambda_+ \tau} + e^{-\lambda_- \tau}\right) + \frac{2 - \lambda_+ - \lambda_-}{(\lambda_+ - \lambda_-)}\left(e^{-\lambda_- \tau} - e^{-\lambda_+ \tau}\right)\right\}\Delta T(0) + \\ & + \frac{S}{l\kappa(\lambda_+ - \lambda_-)}\left(e^{-\lambda_- \tau} - e^{-\lambda_+ \tau}\right)\Delta N(0)\,,\end{aligned}\tag{S-53}$$

in which $\tau = t/\tau_0$ is the time measured in units of $\tau_0 = \kappa/g$ and $\lambda_\pm$ are the eigenvalues of the matrix appearing in Eq. (10) in the main text

$$\lambda_\pm = \frac{1}{2}\left(1 + \frac{L + S^2}{l}\right) \pm \sqrt{\frac{1}{4}\left(1 + \frac{L + S^2}{l}\right)^2 - \frac{L}{l}}\,.\tag{S-54}$$

If thermoelectric effects are neglected, we find $\lambda_+ = 1$ and $\lambda_- = K_\mathrm{r} L_\mathrm{LS}/L_0$, with $L_\mathrm{LS}$ given in Eq. (5) in the main text. These two eigenvalues define different time scales for the equilibration of particle with respect to temperature imbalances. In the main text we will be interested in the time evolution of relative imbalances $\Delta N(t)/N$ and $\Delta T(t)/T$, in which $N$ is the total number of particles in the system and $T$ the average temperature between the two reservoirs. It is interesting to study the adaption of Eq. (S-53) to these quantities

$$\begin{aligned}\frac{\Delta N(\tau)}{N} =& \frac{1}{2}\left\{\left(e^{-\lambda_+ \tau}+e^{-\lambda_- \tau}\right)-\frac{2-\lambda_+-\lambda_-}{(\lambda_+-\lambda_-)}\left(e^{-\lambda_- \tau}-e^{-\lambda_+ \tau}\right)\right\}\frac{\Delta N(0)}{N}+ \\ & +\frac{\kappa S T}{N}\frac{\left(e^{-\lambda_- \tau}-e^{-\lambda_+ \tau}\right)}{\lambda_+-\lambda_-}\frac{\Delta T(0)}{T}, \\ \frac{\Delta T(\tau)}{T} =& \frac{1}{2}\left\{\left(e^{-\lambda_+ \tau}+e^{-\lambda_- \tau}\right)+\frac{2-\lambda_+-\lambda_-}{(\lambda_+-\lambda_-)}\left(e^{-\lambda_- \tau}-e^{-\lambda_+ \tau}\right)\right\}\frac{\Delta T(0)}{T}+ \\ & +\frac{SN}{l\kappa T}\frac{\left(e^{-\lambda_- \tau}-e^{-\lambda_+ \tau}\right)}{(\lambda_+-\lambda_-)}\frac{\Delta N(0)}{N}.\end{aligned} \quad (\text{S-55})$$

Considering Eq. (11) in the main text, giving the explicit expression of $S$, we notice that the proportionality factor between $\Delta N(\tau)/N$ and the initial temperature imbalance $\Delta T(0)/T$ is essentially governed by the ratio $(T/T_F)^2$ which could be as low as $10^{-2}$ in cold atom clouds. The proportionality factor between $\Delta T(\tau)/T$ and $\Delta N(0)/N$ depends on temperature only through $L_{\text{LS}}$, which has a much weaker dependence on temperature. This explains why the time evolution of the relative temperature imbalance $\Delta T(t)/T$ is much more suitable to study thermoelectric effects in interacting 1D clouds.

## S-V. MEAN FREE PATH OF ENERGY MODES

In this section, the mean free path Eq. (14) in the main text is derived. We follow the discussion in Ref. [1] concerning the localization of the energy modes in presence of a smooth disorder and extend it to the bosonic case.

For a homogeneous system, the eigenstates of Eq. (S-23) are plane-waves $\phi_\omega(x)=e^{i\omega x/u}/\sqrt{L}$, in which $L$ is the system size. Small disorder fluctuations $\delta\rho(x)\ll n_0$ lead to a perturbation $\delta h$ to the eigenstate equation Eq. (S-23), $h_0\to h_0+\delta h$. The effects on $h_0$ of density fluctuations can be readily understood by making explicit the boson density $n_0$ through the set of equalities in Eq. (S-7). The ratio $u/K$ as a function of density is known obtained both in the strongly and weakly interacting limit [6] by comparing the hydrodynamic solution with the exact solution for a system of delta interacting bosons, the Lieb-Liniger model [7]. Defining the dimensionless parameter $\gamma=mg/\hbar^2 n_0$, in which $g>0$ is the interaction strength,

$$\begin{aligned}\frac{u}{K} &= \frac{\pi\hbar}{m}n_0\left(1-\frac{8}{\gamma}\right), & \text{for } \gamma\gg 1,\\ \frac{u}{K} &= \frac{\pi\hbar}{m}n_0\frac{\gamma}{\pi^2}\left(1-\frac{\sqrt{\gamma}}{2\pi}\right), & \text{for } \gamma\ll 1.\end{aligned} \quad (\text{S-56})$$

Making the substitution $n_0(x)=n_0+\delta\rho(x)$, the perturbation $\delta h$ is derived to linear order in $\delta\rho$

$$\begin{aligned}\delta h &= -\frac{1}{2}\left(\frac{\pi\hbar}{m}\right)^2 n_0\left(1-\frac{8}{\gamma}\right)\left[\delta\rho\frac{\partial^2}{\partial x^2}+\frac{\partial^2}{\partial x^2}\delta\rho+2\frac{\partial}{\partial x}\delta\rho\frac{\partial}{\partial x}\right]+\frac{8}{\gamma}n_0\left(\frac{\pi\hbar}{m}\right)^2\frac{\partial}{\partial x}\delta\rho\frac{\partial}{\partial x}, & \gamma\gg 1,\\ \delta h &= -\frac{1}{2}\left(\frac{\pi\hbar}{m}\right)^2 n_0\frac{\gamma}{\pi^2}\left(1-\frac{\sqrt{\gamma}}{2\pi}\right)\left[\delta\rho\frac{\partial^2}{\partial x^2}+\frac{\partial^2}{\partial x^2}\delta\rho+2\frac{\partial}{\partial x}\delta\rho\frac{\partial}{\partial x}\right]+n_0\left(\frac{\pi\hbar}{m}\right)^2\frac{\gamma}{\pi^2}\left(1-\frac{3\sqrt{\gamma}}{4\pi}\right)\frac{\partial}{\partial x}\delta\rho\frac{\partial}{\partial x}, & \gamma\ll 1.\end{aligned} \quad (\text{S-57})$$

The perturbation $\delta h$ gives a lifetime $\tau(\omega)$ to the eigenstates $\phi_\omega(x)$, which is obtained applying Fermi's golden rule

$$\frac{1}{\tau(\omega)}=\frac{\pi}{\omega}\sum_{\omega'\neq\omega}\overline{|\langle\phi_{\omega'}|\delta h|\phi_\omega\rangle|^2}\delta(\omega^2-\omega'^2). \quad (\text{S-58})$$

The lifetime $\tau(\omega)$ also defines the mean free path through the simple equality $\tau(\omega)=\xi(\omega)/u$. The overline $\overline{\langle\cdot\rangle}$ defines the average over the disorder fluctuations $\delta\rho$. The matrix element $\langle\phi_{\omega'}|\delta h|\phi_\omega\rangle$ involves integrals of the form

$$\int dx\,\phi^*_{\omega'}(x)\left[\delta\rho\frac{\partial^2}{\partial x^2}+\frac{\partial^2}{\partial x^2}\delta\rho+2\frac{\partial}{\partial x}\delta\rho\frac{\partial}{\partial x}\right]\phi_\omega(x)$$

which are obtained from the first contributions in Eq. (S-57). The energy conservation condition in Eq. (S-58) requires $\omega'=-\omega$, implying $\phi^*_{\omega'}(x)=\phi_\omega(x)$, and the above integral is then zero. Then, only the last terms in Eq.

(S-57) provide non-zero contributions to $\langle\phi_{\omega'}|\delta h|\phi_\omega\rangle$, which reads

$$\langle\phi_{\omega'}|\delta h|\phi_\omega\rangle = \frac{A}{L}\left(\frac{\omega}{u}\right)^2 \int dx\, \delta\rho(x)\, e^{2i\omega x/u}\,, \tag{S-59}$$

in which the prefactor $A$ assumes different forms in the strongly and weakly interacting limit

$$\begin{aligned} A &= \frac{8}{\gamma} n_0 \left(\frac{\pi\hbar}{m}\right)^2, & \gamma \gg 1\,, \\ A &= n_0\, \gamma \left(\frac{\hbar}{m}\right)^2 \left(1 - \frac{3\sqrt{\gamma}}{4\pi}\right), & \gamma \ll 1\,. \end{aligned} \tag{S-60}$$

Substituting Eq. (S-59) into Eq. (S-58), one finds

$$\frac{1}{\xi(\omega)} = \frac{\pi}{u\omega}\frac{A^2}{L^2}\left(\frac{\omega}{u}\right)^4 \int dxdx' K(x-x') e^{2i\omega(x-x')/u} \sum_{\omega'\neq\omega} \delta(\omega^2 - \omega'^2)\,. \tag{S-61}$$

In this expression the average over disorder has been carried out and $K(x-x') = \langle\delta\rho(x)\delta\rho(x')\rangle/\langle\delta\rho(x)^2\rangle$ is the disorder correlation function. Taking the $L\to\infty$ limit one obtains

$$\frac{1}{\xi(\omega)} = \frac{\omega^2 A^2 \langle\delta\rho^2\rangle}{4u^6} \int_{-\infty}^{\infty} dx K(x) e^{2i\omega x/u}\,. \tag{S-62}$$

Taking a Gaussian form for the correlation function $K(x) = e^{-x^2/l_D^2}$, in which $l_D$ is the disorder correlation length, the mean free path Eq. (14) in the main text is derived.